# The effect of publishing a highly cited paper on journal's impact factor: a case study of the Review of Particle Physics

**Forthcoming in *Learned Publishing***

Weishu Liu

ORCID: 0000-0001-8780-6709

wsliu08@163.com

School of Information Management and Engineering, Zhejiang University of Finance and Economics, Hangzhou 310018, Zhejiang, China

Fang Liu

ORCID: 0000-0001-7337-8694

liu_fang2014@163.com

School of Accounting, Zhejiang University of Finance and Economics, Hangzhou 310018, Zhejiang, China

Chao Zuo

ORCID: 0000-0002-5177-5813

chaozuo1982@gmail.com

School of Management and E-Business, Zhejiang Gongshang University, Hangzhou 310018, Zhejiang, China

Junwen Zhu

ORCID: 0000-0002-3543-674X

jwzhu@sjtu.edu.cn

Corresponding author

Graduate School of Education, Shanghai Jiao Tong University, Shanghai 200240, China

**Abstract**: A single highly cited article can give a big but temporary lift in its host journal's impact factor evidenced by the striking example of "A short history of SHELX" published in Acta Crystallographica Section A. By using Journal Citation Reports and Web of Science's citation analysis tool, we find a more general and continuous form of this phenomenon in the Particle Physics field. The highly-cited "Review of Particle Physics" series have been published in one of the major Particle Physics journals biennially. This study analyses the effect of these articles on the Impact Factor (IF) of the host journals. The results show that the publication of Review of Particle Physics articles has a direct effect of lifting the IF of its host journal. However the effect on the IF varies according to whether the host journal already has a relatively high or low IF, and the number of articles that it publishes. The impact of these highly cited articles clearly demonstrates the limitations of journal impact factor, and endorses the need to use it more wisely when deciding where to publish and how to evaluate the relative impact of a journal.

**Keywords**: Impact factor; Journal Citation Reports; Highly cited paper; Research assessment; Review of Particle Physics

Key points:

- A more general and continuous form of "A short history of SHELX" phenomenon has been found.
- The Review of Particle Physics series, which are called the bible in the Particle Physics field, are highly cited by other research papers.
- The Review of Particle Physics phenomenon is found in different journals and is affected by the host journal's IF and publication size.
- The Review of Particle Physics phenomenon give a big or small, but temporary, lift in its host journal's impact factor.

**Introduction**

The Journal Impact Factor (JIF) was devised by Irving H. Sher and Eugene Garfield to help evaluate and select journals (Garfield, 2006). Although the JIF was originally used as a tool to help librarians purchase journals (Garfield, 1972), this indicator has been widely used and discussed in research and researcher related evaluation (Liu, Hu, & Gu, 2016; Tang, 2013; Waltman, 2016). JIF as one of the most important indicators of a journal's quality/impact has attracted wide attention from scientists, journal editors to publishers (Martin, 2016). Meanwhile, increasing efforts have focused on the undue and inappropriate use of JIF in research and researcher related evaluation (Alberts, 2013; Bornmann, Marx, Gasparyan, & Kitas, 2012; Hicks, Wouters, Waltman, de Rijcke, & Rafols, 2015; Lariviere et al., 2016). One of the main deficiencies of the JIF is that the calculation of JIF is based on the arithmetic mean of a highly skewed distribution of citations (Bornmann & Leydesdorff, 2017; Lariviere et al., 2016).

A few very highly cited papers can boost the host journal's JIF due to the arithmetic mean calculation of JIF. An extreme example is the publishing of "A short history of SHELX" in Acta Crystallographica Section A (Sheldrick, 2008). According to the Web of Science, this paper has been cited more than 53,000 times and is one of the top 10 most cited papers in Web of Science. This single highly cited paper gave a big but temporary lift to its host journal's impact factor (Dimitrov, Kaveri, & Bayry, 2010; Foo, 2013; Krauskopf, 2013). More specifically, the JIF of Acta Crystallographica Section A skyrocketed from 2.051 in 2008 to 49.926 in 2009 and stayed at 54.333 in 2010, but dropped to 2.076 in 2011. This striking example demonstrates a single paper's dramatic but temporary role in lifting the host journal's impact factor which may further bias the whole citation-based evaluation system.

However, the "A short history of SHELX" phenomenon is not alone and this phenomenon can be demonstrated by the "Review of Particle Physics" (RPP) series in the Particle Physics field. The very highly cited RPP series provide "a compendium of experimental data and reviews put out by the Particle Data Group" (Brooks, 2007). The summaries are published biennially in one of the major journals in Particle Physics (Particle Data Group, 2016). Since the RPP series have been published in different journals in recent years, a deep investigation on their effect of JIF can provide a more general and comprehensive view of the "A short history of SHELX" phenomenon.

**Data and method**

The Journal Impact Factor is defined as "all citations to the journal in the current JCR year to items published in the previous two years, divided by the total number of scholarly items (these comprise articles, reviews, and proceedings papers) published in the journal in the previous two years" (Clarivate Analytics, 2016). For example, journal A's 2015 impact factor can be calculated by the following equation (1):

$$\text{Equation (1)} \quad \text{Journal A's 2015 impact factor} = \frac{\text{Citations in 2015 to all items published by Journal A in 2013} - 2014}{\text{Total number of scholarly items published by Journal A in 2013} - 2014}$$

In this study, we want to probe the lifting effect of "RPP" on the host journal's impact factor. We recalculated the impact factor by excluding the citations attracted by this highly cited review as shown in equation (2). Accordingly, this review was also be excluded from the counting of the number of scholarly items.

$$\text{Equation (2)} \quad \text{Journal A's 2015 adjusted impact factor} = \frac{\text{Citations in 2015 to all items(Excluding "RPP") published by Journal A in 2013} - 2014}{\text{Total number of scholarly items (Excluding "RPP") published by Journal A in 2013} - 2014}$$

**Analyses**

**The Review of Particle Physics effect**

The "Review of Particle Physics" has been called the bible in Particle Physics field. These biennially updated reviews have attracted large number of citations from the scientific community and may have an unexpected lifting effect on the host journal's impact factor. Eight recently published "Review of Particle Physics" papers in four different journals during 2002 to 2016 are selected for this study. Each journal has published two of these eight reviews. These four journals have different levels of impact factor and publishing volume and are also published by four different publishers. These journal level characteristics provide an opportunity to probe the lifting effect of a single highly cited paper on the host journal's impact factor.

Table 1 provides the detailed information of these eight recently published reviews. The four journals are: Physical Review D, Physics Letters B, Journal of Physics G-Nuclear and Particle Physics, and Chinese Physics C. Using Web of Science's citation analysis tool, we identified the RPP's total number of citations, and the number of citations received in the 2nd and 3rd years. The data were collected from the library of Xi'an Jiao Tong University on 23 March 2017. All these RPPs have been cited more than 2,000 times except the latest one published in 2016 which has had limited time to attract citations. A large proportion of the total citations of the RPPs occurs in the 2nd and 3rd year after publishing, and so may have a significant effect on the host journal's impact factor.

The numbers of citations of each RPP received during the second and third year after publishing are relatively stable as shown in Table 1. In order to probe the lifting effect of RPP on host journal's impact factor, the host journal's actual impact factor and publishing volume should also be considered. In the following sections, we try to probe the lifting effect on impact factor of four host journals respectively.

[INSERT TABLE 1 HERE]

**Physical Review D: high IF, large publishing volume**

Hagiwara et al. (2002) and Beringer et al. (2012) are the two RPPs that were published in Physical Review D. According to the Journal Citation Reports, this journal belongs to two Web of Science categories: Astronomy & Astrophysics and Physics, Particles & Fields. The JIFs of this journal during the past 15 years have always been high (ranging from 4.3 to 5.2 with the median value of 4.69). The JIF percentiles of this journal during all the study period were always higher than 70%, that is to say, this journal was almost always a Quartile 1 (Q1) journal. This journal also has a large yearly publishing volume of 2,000 or more citable items (i.e., articles and reviews) every year. To sum up, the Physical Review D is a journal with high impact factor and large yearly publishing volume.

Figure 1 depicts the dynamics of Physical Review D's impact factors during 2001 to 2015 as shown by the blue line. In order to probe the effect of RPPs on the journal's impact factor, we recalculated the impact factor of Physical Review D by excluding the citations attracted by these specific reviews. The red line shows the changing trends of Physical Review D's adjusted impact factors.

The values of Physical Review D's impact factor in 2003, 2004, 2013, and 2014 were only a bit higher than its values of adjusted impact factor in these years from both absolute and relative perspectives. As anticipated, a single highly cited paper's lifting effect on the host journal's impact factor was diluted by the host journal's high impact factor and large yearly publishing volume. Therefore the "Review of Particle Physics" effect on this type of journal with high impact factor and large annual publishing volume is not that remarkable from both absolute and relative perspectives.

[INSERT FIGURE 1 HERE]

**Physics Letters B: high IF, medium publishing volume**

Another two RPP papers were published in Physics Letters B in 2004 and 2008 respectively (Amsler et al., 2008; Eidelman et al., 2004). Similar to Physical Review D, Physics Letters B is also a high impact factor journal. According to Journal Citation Reports, this journal is a Quartile 1 (Q1) journal during the whole study period. However, this journal only publishes about 1,000 citable items (i.e., articles and reviews) annually, which is much lower than the annual publishing volume of the previous Physical Review D.

Figure 2 depicts the dynamics of Physics Letters B's impact factors and adjusted impact factors during 2001 to 2015 as shown by the blue line and red line respectively. As shown by the blue line, three peak periods of Physics Letters B's impact factor can be identified. Specifically, three peak periods are 2005-2006, 2009-2010, and 2013-2014. The red line also illustrates the adjusted impact factor of Physics Letters B by excluding the "Review of Particle Physics" effect. The adjusted impact factor of Physics Letters B was much more stable before 2013. The values of Physics Letters B's impact factor increased significantly in 2005, 2006, 2009, and 2010 due to the RPP effect evidenced by the gaps between the blue and red lines. Compared to Physical Review D, the "Review of Particle Physics" effect is more significant especially from the absolute perspective. This may due to the relatively smaller publishing volume of Physics Letters B than Physical Review D.

Surprisingly, another higher peak of Physics Letters B's impact factor appeared in 2013 and 2014. However, this peak cannot be attributed to "Review of Particle Physics". By investigating papers published in Physics Letters B during 2011 and 2013, we find another two highly cited papers published in Physics Letters B in 2012 (Aad et al., 2012; Chatrchyan et al., 2012). Both these two papers have been cited about 4,000 times so far. The combined effect of these two papers leads the impact factor of Physics Letters B to a higher level in 2013 and 2014.

[INSERT FIGURE 2 HERE]

**Journal of Physics G-Nuclear and Particle Physics: medium IF, small publishing volume**

The 2006 and 2010 RPP articles were published in Journal of Physics G-Nuclear and Particle Physics (Nakamura et al., 2010; Yao et al., 2006). Different from the previous two high impact factor journals, the JIF of this journal is relative lower. According to Journal Citation Reports, the values of Journal of Physics G's impact factor fluctuate around 2 with a few peaks. This journal is allocated to two Web of Science categories: Physics, Nuclear and Physics, Particles & Fields. This journal was always a Quartile 2 (Q2) or Quartile 3 (Q3) except during two peak periods. This journal only publishes about 200 citable items every year. To sum up, we can regard Journal of Physics G-Nuclear and Particle Physics as a journal with medium impact factor and small publishing volume.

Figure 3 depicts the dynamics of Journal of Physics G's impact factors and adjusted impact factors during 2001 to 2015 as shown by the blue line and red line respectively. As shown by the blue line, two peak periods of Journal of Physics G's impact factor can be clearly identified. Specifically, two peak periods are 2007-2008 and 2011-2012 which happens to be the two subsequent years after the publishing of "Review of Particle Physics" in Journal of Physics G. The red line also illustrates the adjusted impact factor of Journal of Physics G by excluding the "Review of Particle Physics" effect. Different from the actual impact factor, the adjusted impact factor of Journal of Physics G is much more stable. The values of Journal of Physics G's impact factor increased significantly in 2007, 2008, 2011, and 2012 via the "Review of Particle Physics" effect evidenced by the gaps between the blue and red lines. Compared to the previous two journals, the "Review of Particle Physics" effect is more significant on Journal of Physics G from both absolute and relative perspectives. According to the definition of JIF, this is due to this journal's relatively lower impact factor and smaller yearly publishing volume.

[INSERT FIGURE 3 HERE]

**Chinese Physics C: low IF, small publishing volume**

Interestingly, two latest RPP papers were published by Chinese Physical Society in the journal of "Chinese Physics C" (Olive et al., 2014; Patrignani et al., 2016). This journal's current title was changed from "High Energy Physics and Nuclear Physics-Chinese Edition" in 2008. According to Journal Citation Reports, both these two titles suffered from low impact factor before 2013. These two titles were always Quartile 4 (Q4) journals in both Physics, Nuclear and Physics, Particles & Fields categories before 2013. Although the Chinese Physics C was a low impact factor journal, this journal has successfully bid on publishing the RPP (Barnett, 2014). Apart from having enough funding for publishing, the choice of a Chinese journal to publish this classical review may partly due to the rapid rising of China's scientific research (Liu, Hu, Tang, & Wang, 2015; Zhou & Leydesdorff, 2006) and growing global interests in China (Liu, Tang, Gu, & Hu, 2015). Apart from low impact factor, only about 200 citable items are published by this journal annually. That is to say, Chinese Physics C is a small size and low impact factor journal.

Figure 4 depicts the dynamics of Chinese Physics C's impact factor and adjusted impact factor as shown by the blue line and red line respectively. Since the change of journal title in 2008, we have only taken the period of 2009-2015 into consideration. By referring to the blue line, this journal's actual impact factor began to rise from 2013. Chinese Physics C's impact factor skyrocketed from 1.313 in 2014 to 3.761 in 2015, and the red line reveals that this was due to the effect of publishing the 2014 RPP article. The significant increasing of Chinese Physics C's impact factor is due to this journal's low impact factor and small yearly publishing volume.

Since two latest reviews were published in this specific journal in 2014 and 2016 respectively, this journal's impact factor will very likely maintain high level in the following three years. That is to say, we can predict that this journal's impact factors in 2015, 2016, and 2017 are very likely to be high.

[INSERT FIGURE 4 HERE]

**Conclusion**

By using the Journal Citation Reports and Web of Science's citation analysis tool, this study finds the existence of a "Review of Particle Physics" phenomenon in the Particle Physics field which is a more general and continuous form of the "A short history of SHELX" phenomenon in Acta Crystallographica Section A. A single highly cited article can lift its host journal's impact factor temporarily. However, the lifting effect on JIF is influenced by the host journal's impact factor and yearly publishing volume. The JIFs of high impact factor and large size journals can only be influenced by the RPP effect marginally. However, low impact factor and small size journals' impact factors can be lifted greatly from both absolute and relative perspectives. The "A short history of SHELX" phenomenon belongs to this case.

This unique case study gives us a real-world example about the limitations of JIF once again (Amin & Mabe, 2000; Bornmann, 2017; Lariviere et al., 2016; Liu, Liu, Zuo & Zhu, 2017). The citation distribution for articles in a specific journal may be highly skewed. Any journal which gets the opportunity to publish the "Review of Particle Physics" can get a big or small, but temporary, lift in its impact factor as a by-product. Since the effect of RPP is temporary, a researcher who seeks to publish in a high JIF journal may submit a manuscript to a host journal when its JIF is at its temporarily high point, however, the manuscript may be published when this journal's JIF returns to normal. Similarly, the opposite situation may also happen.

The wide use of JIF in research and researcher evaluation has become a big concern for many stakeholders. However, the phenomenon found in this study demonstrates the ability to distort this journal-based metric. Although JIF is a quite useful indicator and the cases mentioned in this study are also special, this widely used indicator should be used more wisely especially in Particle Physics field. The JIF may be used with other indicators such as geometric journal impact factors (Thelwall & Fairclough, 2015) and H index (Liu, 2017).

**References**

Aad, G., Abajyan, T., Abbott, B., Abdallah, J., Khalek, S. A., Abdelalim, A. A., et al. (2012). Observation of a new particle in the search for the Standard Model Higgs boson with the ATLAS detector at the LHC. *Physics Letters B, 716*(1), 1-29. doi: 10.1016/j.physletb.2012.08.020

Alberts, B. (2013). Impact Factor Distortions. *Science, 340*(6134), 787-787. doi: 10.1126/science.1240319

Amin, M., & Mabe, M. (2000). Impact factors:use and abuse. *Perspectives in Publishing, 1, 1-6.*

Amsler, C., Doser, M., Antonelli, M., Asner, D. A., Babu, K. S., Baer, H., et al. (2008). Review of particle physics. *Physics Letters B, 667*(1-5), 1-+. doi: 10.1016/j.physletb.2008.07.018

Barnett, M. (2014). Particle Data Group 57 years of service. https://science.energy.gov/~/media/hep/hepap/pdf/20141208/barnett_hepap_12_14b.pdf Accessed on October 18, 2017.

Beringer, J., Arguin, J. F., Barnett, R. M., Copic, K., Dahl, O., Groom, D. E., et al. (2012). Review of particle physics. *Physical Review D, 86*(1), 1504. doi: 10.1103/PhysRevD.86.010001

Bornmann, L. (2017). Confidence intervals for Journal Impact Factors. *Scientometrics*, 111(3), 1869-1871. doi: 10.1007/s11192-017-2365-3

Bornmann, L., & Leydesdorff, L. (2017). Skewness of citation impact data and covariates of citation distributions: A large-scale empirical analysis based on Web of Science data. *Journal of Informetrics, 11*(1), 164-175. doi:10.1016/j.joi.2016.12.001

Bornmann, L., Marx, W., Gasparyan, A. Y., & Kitas, G. D. (2012). Diversity, value and limitations of the journal impact factor and alternative metrics. *Rheumatology International, 32*(7), 1861-1867. doi: 10.1007/s00296-011-2276-1

Brooks, T. (2007). The hottest citation. http://www.symmetrymagazine.org/article/april-2007/the-hottest-citation. Accessed on 22 April 2017.

Clarivate Analytics. (2016). Journal Citation Reports. http://ipscience-help.thomsonreuters.com/incitesLiveJCR/JCRGroup/jcrOverview.html. Accessed on 22 April 2017.

Chatrchyan, S., Khachatryan, V., Sirunyan, A. M., Tumasyan, A., Adam, W., Aguilo, E., et al. (2012). Observation of a new boson at a mass of 125 GeV with the CMS experiment at the LHC. *Physics Letters B, 716*(1), 30-61. doi: 10.1016/j.physletb.2012.08.021

Dimitrov, J. D., Kaveri, S. V., & Bayry, J. (2010). Metrics: journal's impact factor skewed by a single paper. *Nature, 466*(7303), 179-179. doi:10.1038/466179b

Eidelman, S., Hayes, K. G., Olive, K. A., Aguilar-Benitez, M., Amsler, C., Asner, D., et al. (2004). Review of particle physics. *Physics Letters B, 592*(1-4), 1-1109. doi: 10.1016/j.physletb.2004.06.001

Foo, J. Y. A. (2013). Implications of a Single Highly Cited Article on a Journal and Its Citation Indexes: A Tale of Two Journals. *Accountability in Research-Policies and Quality Assurance, 20*(2), 93-106. doi: 10.1080/08989621.2013.767124

Garfield, E. (1972). Citation Analysis as a Tool in Journal Evaluation. *Science, 178*(4060), 471-479. doi: 10.1126/science.178.4060.471

Garfield, E. (2006). The history and meaning of the journal impact factor. *JAMA, 295*(1), 90-93. doi: 10.1001/jama.295.1.90

Hagiwara, K., Hikasa, K., Nakamura, K., Tanabashi, M., Aguilar-Benitez, M., Amsler, C., et al. (2002). Review of particle physics. *Physical Review D, 66*(1), 985. doi: 10.1103/PhysRevD.66.010001

Hicks, D., Wouters, P., Waltman, L., de Rijcke, S., & Rafols, I. (2015). The Leiden Manifesto for research metrics. *Nature, 520*(7548), 429-431. doi:10.1038/520429a

Krauskopf, E. (2013). Deceiving the Research Community Through Manipulation of the Impact Factor. *Journal of the American Society for Information Science and Technology, 64*(11), 2403-2403. doi: 10.1002/asi.22905

Lariviere, V., Kiermer, V., MacCallum, C. J., McNutt, M., Patterson, M., Pulverer, B., et al. (2016). A simple proposal for the publication of journal citation distributions. *bioRxiv*. doi: 10.1101/062109

Liu, W. (2017). The changing role of non-English papers in scholarly communication: Evidence from Web of Science's three journal citation indexes. *Learned Publishing*, 30(2), 115-123. doi: 10.1002/leap.1089

Liu, W., Hu, G., & Gu, M. (2016). The probability of publishing in first-quartile journals. *Scientometrics, 106*(3), 1273-1276. doi: 10.1007/s11192-015-1821-1

Liu, W., Hu, G., Tang, L., & Wang, Y. (2015). China's global growth in social science research: Uncovering evidence from bibliometric analyses of SSCI publications (1978–2013). *Journal of Informetrics, 9*(3), 555-569. doi: 10.1016/j.joi.2015.05.007

Liu, F., Hu, G., Tang, L., & Liu, W. (2017). The penalty of containing more non-English articles. *Scientometrics*, in press. doi: 10.1007/s11192-017-2577-6

Liu, W., Tang, L., Gu, M., & Hu, G. (2015). Feature report on China: a bibliometric analysis of China-related articles. *Scientometrics, 102*(1), 503-517. doi: 10.1007/s11192-014-1371-y

Martin, B. R. (2016). Editors' JIF-boosting stratagems – Which are appropriate and which not? *Research Policy*, 45(1), 1-7. doi: 10.1016/j.respol.2015.09.001

Nakamura, K., Hagiwara, K., Hikasa, K., Murayama, H., Tanabashi, M., Watari, T., et al. (2010). Review of particle physics. *Journal of Physics G-Nuclear and Particle Physics, 37*(7A), 1-5. doi: 10.1088/0954-3899/37/7a/075021

Olive, K. A., Agashe, K., Amsler, C., Antonelli, M., Arguin, J. F., Asner, D. M., et al. (2014). Review of particle physics. *Chinese Physics C, 38*(9), 1658. doi: 10.1088/1674-1137/38/9/090001

Particle Data Group. (2016). What the Particle Data Group is. http://pdg.lbl.gov/2016/html/what_is_pdg.html. Accessed on 22 April 2017.

Patrignani, C., Agashe, K., Aielli, G., Amsler, C., Antonelli, M., Asner, D. M., et al. (2016). Review of particle physics. *Chinese Physics C, 40*(10), 1790. doi: 10.1088/1674-1137/40/10/100001

Sheldrick, G. M. (2008). A short history of SHELX. *Acta Crystallographica Section A, 64*, 112-122. doi: 10.1107/s0108767307043930

Thelwall, M., & Fairclough, R. (2015). Geometric journal impact factors correcting for individual highly cited articles. *Journal of Informetrics*, 9(2), 263-272. doi: 10.1016/j.joi.2015.02.004

Tang, L. (2013). Does "birds of a feather flock together" matter-Evidence from a longitudinal study on US-China scientific collaboration. *Journal of Informetrics, 7*(2), 330-344. doi: 10.1016/j.joi.2012.11.010

Waltman, L. (2016). A review of the literature on citation impact indicators. *Journal of Informetrics*, 10(2), 365-391. doi: 10.1016/j.joi.2016.02.007

Yao, W. M., Amsler, C., Asner, D., Barnett, R. M., Beringer, J., Burchat, P. R., et al. (2006). Review of particle physics. *Journal of Physics G-Nuclear and Particle Physics, 33*(1), 1-+. doi: 10.1088/0954-3899/33/1/001

Zhou, P., & Leydesdorff, L. (2006). The emergence of China as a leading nation in science. *Research Policy*, 35(1), 83-104. doi: 10.1016/j.respol.2005.08.006

Table 1 Summary of "Review of Particle Physics" article publications

| Authors (Year) | Journal Title | Publisher | Second Year Citations | Third Year Citations | Total Citations |
|---|---|---|---|---|---|
| Hagiwara et al. (2002) | Physical Review D | American Physical Society, USA | 873 | 1100 | 2887 |
| Eidelman et al. (2004) | Physics Letters B | Elsevier Science B.V., Netherlands | 1188 | 1292 | 3707 |
| Yao et al. (2006) | Journal of Physics G-Nuclear and Particle Physics | IOP Publishing Ltd, UK | 1135 | 1494 | 3854 |
| Amsler et al. (2008) | Physics Letters B | Elsevier Science B.V., Netherlands | 1293 | 1702 | 4314 |
| Nakamura et al. (2010) | Journal of Physics G-Nuclear and Particle Physics | IOP Publishing Ltd, UK | 1246 | 1747 | 4117 |
| Beringer et al. (2012) | Physical Review D | American Physical Society, USA | 1528 | 1835 | 4863 |
| Olive et al. (2014) | Chinese Physics C | Chinese Physical Society, China | 1424 | 2015 | 3605 |
| Patrignani et al. (2016) | Chinese Physics C | Chinese Physical Society, China | 24 | n/a | 32 |

Data source: Web of Science Core Collection

Second year citations means number of citations received in the second year after publishing from items in Web of Science Core Collection.

Third year citations means number of citations received in the third year after publishing from items in Web of Science Core Collection.

The data were collected from the library of Xi'an Jiao Tong University on 23 March 2017

Figure 1 The "Review of Particle Physics" effect on Physical Review D

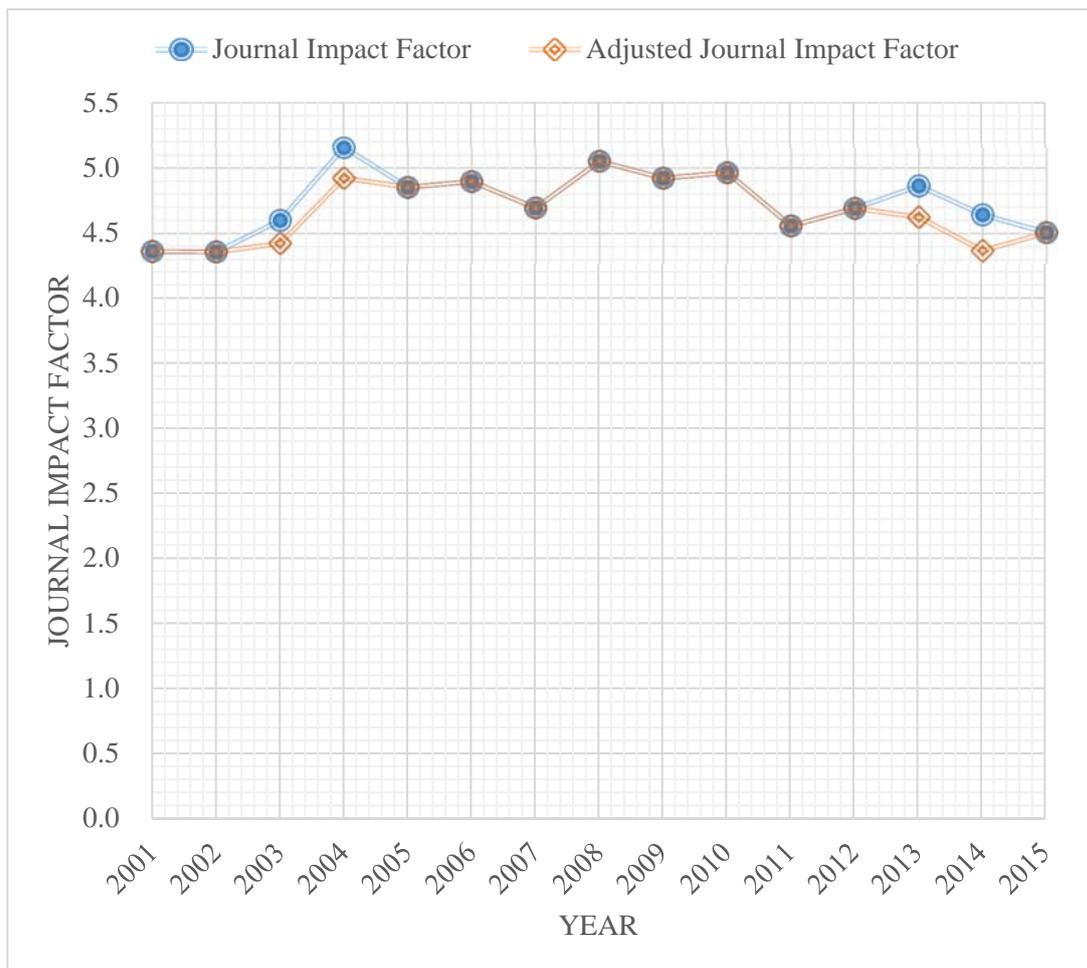

Adjusted journal impact factor, journal impact factor calculated by excluding the citations attracted by this highly cited review. Accordingly, this review should also be excluded from the counting of the number of scholarly items. Data accessed from the library of Xi'an Jiao Tong University on 23 March 2017.

Figure 2 The "Review of Particle Physics" effect on Physics Letters B

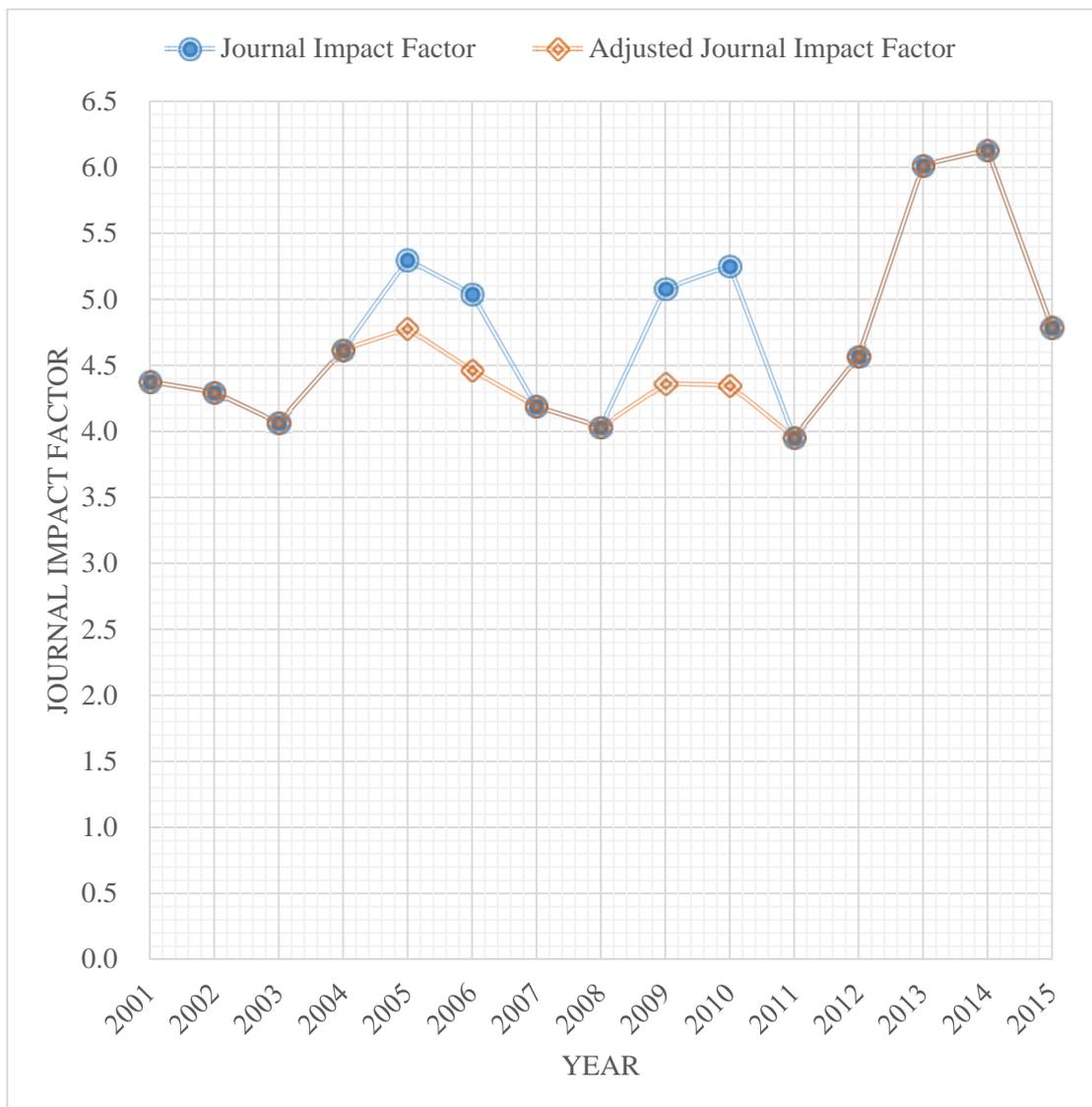

Adjusted journal impact factor, journal impact factor calculated by excluding the citations attracted by this highly cited review. Accordingly, this review should also be excluded from the counting of the number of scholarly items. Data accessed from the library of Xi'an Jiao Tong University on 23 March 2017.

Figure 3 The "Review of Particle Physics" effect on Journal of Physics G

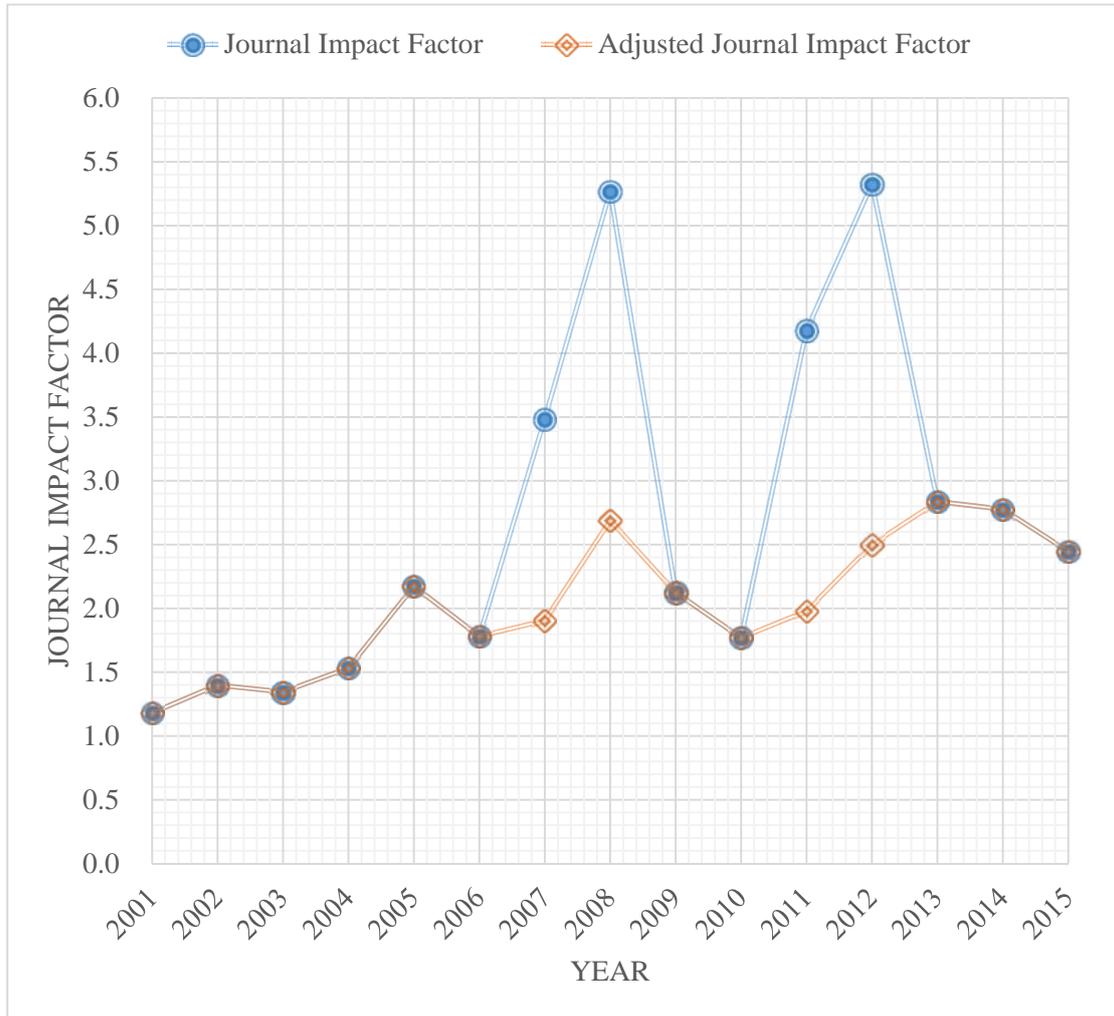

Adjusted journal impact factor, journal impact factor calculated by excluding the citations attracted by this highly cited review. Accordingly, this review should also be excluded from the counting of the number of scholarly items. Data accessed from the library of Xi'an Jiao Tong University on 23 March 2017.

Figure 4 The "Review of Particle Physics" effect on Chinese Physics C

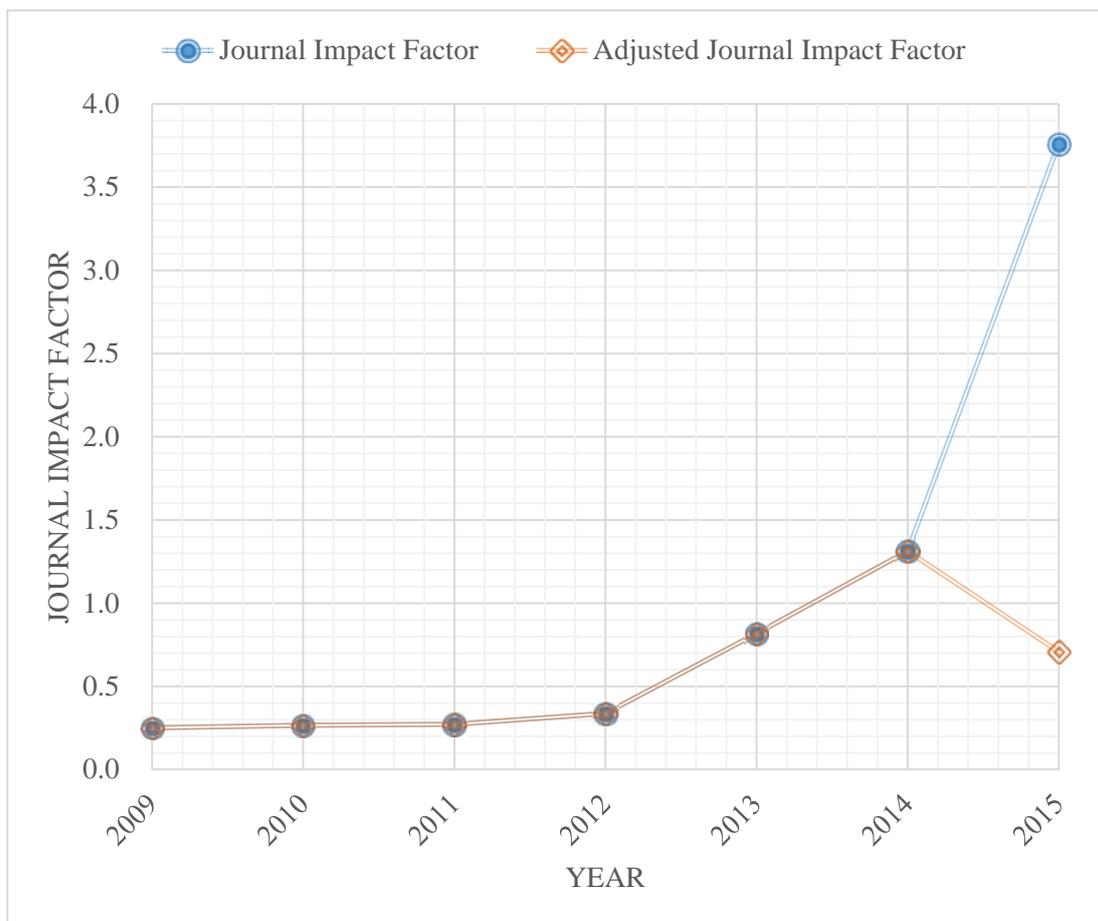

Adjusted journal impact factor, journal impact factor calculated by excluding the citations attracted by this highly cited review. Accordingly, this review should also be excluded from the counting of the number of scholarly items. Data accessed from the library of Xi'an Jiao Tong University on 23 March 2017.